\documentclass[11pt]{article}

\usepackage{mystyle-new}
\usepackage{epsfig,amsmath}
\usepackage{cite}
\usepackage{hyperref}
\usepackage{color}
\usepackage{graphicx}
\definecolor{red}{rgb}{1,0,0}
\def\lesssim{\ \hbox{\raise 2pt \hbox{$<$} \kern -13pt
                     \lower 3pt \hbox{$\sim$}}\ }
\def\greatersim{\ \hbox{\raise 2pt \hbox{$>$} \kern -13pt
                     \lower 3pt \hbox{$\sim$}}\ }

\title{
Aspects of BFKL physics at HERA\footnote{Contribution to the book "From the past to the future, The legacy of Lev Lipatov" \cite{LevBook}.}}
\author{H. Jung\\ Deutsches Elektronen-Synchrotron DESY, Germany}
\begin{document}
\begin{titlepage} 
\maketitle
\vspace*{-7.5cm}
\begin{flushright}
DESY-21-193
\end{flushright}
\vspace*{+6.5cm}

\begin{abstract}
 Aspects of small $x$ physics at HERA are discussed in honor of the work of Lev Lipatov,
who influenced and motivated a whole generation of physicists to investigate this new region of phase space.
The introduction of transverse momentum dependent parton densities is one of the essential aspects
of small $x$ and BFKL physics, and its extension to all flavors with application to various processes 
marks a new milestone.
\end{abstract}

\end{titlepage}

\def\cascade{{\sc Cascade}}
\def\pythia{{\sc Pythia}}
\def\herwig{{\sc Herwig}}
\def\epjpsi{{\sc Epjpsi}}
\newcommand{\ccfm}{Ciafaloni:1987ur,Catani:1989yc,Catani:1989sg,Marchesini:1994wr}
\newcommand{\bfkl}{Kuraev:1976ge,Kuraev:1977fs,Balitsky:1978ic}
\newcommand{\dglap}{Gribov:1972ri,Lipatov:1974qm,Altarelli:1977zs,Dokshitzer:1977sg}
\renewcommand{\thefootnote}{\fnsymbol{footnote}}
\newcommand{\alphas}{\ensuremath{\alpha_\mathrm{s}}}
\newcommand{\asb}{{\bar \alphas}}
\newcommand{\ktp}{k_{t}^{\prime}}
\def\kt{\ensuremath{k_t}}
\newcommand{\as}{\alpha_\mathrm{s}}


\section{Introduction}\label{sec1.1}

It is a funny story how I became interested in small $x$ and BFKL effects and in the work of Lev Lipatov:
I started to work on the H1 calorimeter during my first postdoc at RWTH Aachen from 1988 -- 1993, 
and we were searching for an appropriate physics process for the energy calibration.
One idea was using the decay electrons from $J/\psi$ production, and I started to write the Monte
Carlo event generator \epjpsi ~\cite{Jung:1992za,epjpsi33}. With this tool at hand we investigated possibilities
to measure the gluon density at small $x$, even extrapolating to a hypothetical LEP-LHC $ep$ collider
at high energies at $\sqrt{s} = 1.26$~TeV \cite{Abraham:1990vj}, where clear deviations from DGLAP~\cite{\dglap}
predictions were expected. These studies raised my interest in  small $x$ physics and established
very close contacts to my Russian colleagues at HERA, resulting in a collaboration, which is still ongoing~\cite{Jung:2006ji,Baranov:2003ex,Baranov:2002ei,Baranov:2002th,Baranov:2000ch,Baranov:1999ma}. In these
early days, I never dreamed that I would have a chance to meet Lev in person. Later when he was often at
DESY in Hamburg we got to know each other, and the greatest highlight in my scientific career was when Lev
came once to my QCD and MC lecture at DESY in 2006 \cite{QCDandColliderPhyiscs2006}. Small $x$ physics 
fascinated a whole generation of scientists, resulting in several workshops (for example the HERA workshops~\cite{hera-workshop-1987,hera-workshop-1991,Ingelman:1996ge}, the Lund Small $x$ workshops~\cite{Andersen:2006pg,Andersen:2003xj,Andersson:2002cf} and also the HERA-LHC workshop series~\cite{Alekhin:2005dx,Alekhin:2005dy,Jung:2009eq}).

\section{Calculation of small \boldmath$x$ processes}
In this section, I introduce a derivation of  the BFKL evolution equation \cite{\bfkl} in an heuristic approach, based on the formulation of the DGLAP equation  as a parton branching evolution.

The DGLAP evolution equation can be written in a form introducing a Sudakov form factor $\Delta_a$~\cite{Ellis:1991qj,Hautmann:2017xtx}:
\begin{eqnarray}
\label{dglap-sudakov-integral}
 xf_a(x,\mu^2)  & = & \Delta_a (  \mu^2  ) \  xf_a(x,\mu^2_0)  \nonumber \\
  & &+  \sum_b \int^{\mu^2}_{\mu^2_0} {{d \mu^{\prime 2} } \over \mu^{\prime 2} } {
{\Delta_a (  \mu^2  )}  \over {\Delta_a (  \mu^{\prime 2}   ) }}\int_x^{z_M} {dz} \;P_{ab} (\as(\mu^{\prime 2})) \;\frac{x}{z}f_b\left({\frac{x}{z}},\mu^{\prime 2}\right)   
\end{eqnarray}
with 
\begin{equation}
\label{sudakov}
 \Delta_a ( \mu^2 ) = 
\exp \left(  -  \sum_b  
\int^{\mu^2}_{\mu^2_0} 
{{d \mu^{\prime 2} } 
\over \mu^{\prime 2} } 
 \int_0^{z_M} dz \  z 
\ P_{ba}(\as(\mu^{\prime 2}) , 
 z ) 
\right) 
  \;\; ,
\end{equation}
where $P_{ab}$ is the regularized splitting function for a process $b \to a + c$. The Sudakov form factor gives the probability for non-resolvable parton branching between the scales  $\mu^2$ and  $ \mu^2_0$.  The parameter $z_M$ separates resolvable from non-resolvable branchings. The Sudakov form factor resums all contributions from non-resolvable branchings as well as all virtual corrections. It is shown explicitly in Refs.~\cite{Hautmann:2017xtx,Hautmann:2017fcj}, that the formulation of eqs.(\ref{dglap-sudakov-integral}) leads exactly to the original DGLAP equation at leading order, next-to-leading and next-to-next-to leading orders in the expansion of the splitting functions, provided  $z_M$ is large enough. Away from $x\to 0$, the singularities $1/(1-z)$ in the splitting function for $g \to g g$ and $q \to q g$  are treated with the Sudakov form factor, the resummation of all contributions cancels against the singularity, a feature which we will also use later when discussing BFKL.
\begin{figure}[htbp] 
   \centering
   \includegraphics[width=10cm]{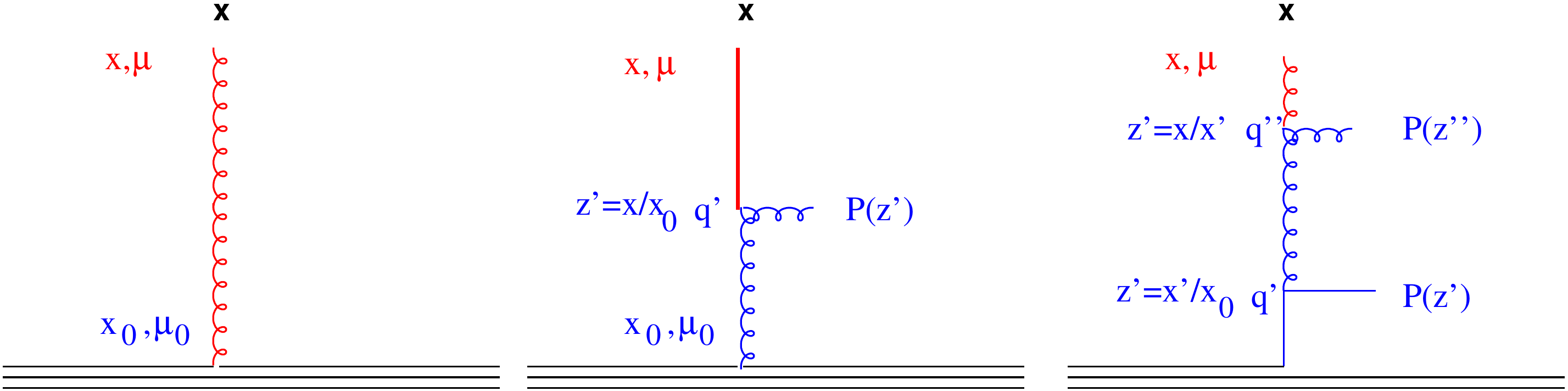}
   \caption{Schematic illustration of the iterative procedure to solve evolution equations: (left) no emission, (middle) one emission, (right) two emissions.}
   \label{Fig:iteration}
\end{figure}
Equation eq.(\ref{dglap-sudakov-integral}) can be solved by iteration, illustrated schematically in Fig.~\ref{Fig:iteration}, leading to 
 an intuitive physical interpretation: the first term of  eq.(\ref{dglap-sudakov-integral}) gives the probability for no emission between the starting scale $\mu_0$ and the scale $\mu$ (left in Fig.~\ref{Fig:iteration}), while the second term describes the branching process (middle and right in Fig.~\ref{Fig:iteration}). 

Although the transverse momenta of the emitted- and the propagating-partons are integrated over in DGLAP, for several applications and in particular for the description in terms of parton showers an exclusive formulation (including transverse momenta) of this equation has advantages.
The evolution equation can be extended for $\kt$ dependent (unintegrated--uPDFs or Transverse Momentum Dependent--TMD) parton densities
${\cal A}(x,{\bf k},t)$ with  ${\bf k}$ being the two-dimensional (transverse) vector of the propagating parton \cite{Hautmann:2017fcj}:
\begin{eqnarray}
\label{TMD}
 x {{\cal A}}_a(x,{\bf k}, \mu^2) 
 &=&  
\Delta_a (  \mu^2  ) \ 
 x{{\cal A}}_a(x,{\bf k},\mu^2_0)  
 + \sum_b 
\int
{{d^2 {\bf q}^{\prime } } 
\over {\pi {\bf q}^{\prime 2} } }
 \ 
{
{\Delta_a (  \mu^2  )} 
 \over 
{\Delta_a (  {\bf q}^{\prime 2})
 ) }
}
 \nonumber\\ 
&& \times
\int_x^{z_M} {dz} \;
P_{ab} (\as) 
\;\frac{x}{z}{ {\cal A}}_b\left(\frac{x}{z}, {\bf k}+(1-z) {\bf q}^\prime , 
{\bf q}^{\prime 2}\right)  
  \;\;  ,     
\end{eqnarray}
 In DGLAP we have:
\begin{equation} 
\label{unintA}
\int  
x \ {\cal A}_a ( x , {\bf k } , \mu^2)  
\  { {d^2 {\bf k }} \over \pi} 
= x {f}_a(x,\mu^2) \; ,
\end{equation}

and the partonic emissions happen in a phase space with increasing scales: $\mu_0 \ll \cdots \ll q' \ll q'' \ll \mu$ for $x_0 \gg \cdots \gg x' \gg x$, as illustrated in Fig.~\ref{Fig:iteration}.

In a formulation, which takes into account color coherence effects, the evolution scale is identified with the angle (angular ordering), while the scale in $\as$ is the transverse momentum of the emitted parton. The angular ordering condition requires for the rescaled transverse momentum $\tilde{q}=q_t/(1-z)$ that $\tilde{q}'' > z \tilde{q}' $, which for $z \to 0$ leads to very weak constraints \cite{\ccfm}.

The BFKL evolution addresses the high energy limit, i.e. $x \to 0$. In this limit, only the gluon contributes and the splitting function becomes  $P_{gg} = 6 \as/z$. The longitudinal momenta become small and  the transverse momenta of the propagating partons cannot be neglected anymore, leading to the introduction of uPDFs. In analogy to the DGLAP case with the singularity at $1/(1-z)$, the singular behavior at $1/z$ can be regulated by resumming all virtual and non-resolvable branchings, described by the so-called {\it non-Sudakov} form factor, which includes now only the small $x$ part of   $P_{gg} $:
\begin{eqnarray}
\Delta_{ns} & = & \exp \left(-\frac{3\alphas}{\pi} \int \frac{dq'^2}{q'^2} \int_z^1\frac{dz'}{z'} 
\Theta(\kt^2 - q'^2) \Theta(q'^2 - \mu_0^2) \right)\\
& = & \exp \left( -\frac{3\alphas}{\pi} \log\frac{1}{z} \log\frac{\kt^2}{\mu_0^2}
\right) \; ,
\end{eqnarray}
 and  the Theta functions limit the available phase space region. While for finite~$z$, the transverse momentum of the emitted parton ${\bf q}$ is related via $\mu = 
| {\bf q} | / (1 - z) $  with the scale $\mu$ of the evolution (angular ordering), at small $z$ this reduces to $\mu = | {\bf q} | $. The total transverse momentum is calculated via  ${\bf k } = - \sum_i {\bf q}_i  $. The angular ordering condition at small $z$ leads to a random walk in $\kt$, since there is no constraint on $q_t$ coming from $\tilde{q}'' > z \tilde{q}' $  for $z \to 0$.

We can rewrite $\Delta_{ns}$ as (using $\asb =\frac{3 \alphas}{\pi})$ :
\begin{eqnarray*}
\Delta_{ns} & = & \exp \left(  \log z \right)^{\asb \log \frac{\kt^2}{\mu_0^2} } \\
 & = & z^\omega \mbox{ with } \omega = \asb \log \frac{\kt^2}{\mu_0^2}
\end{eqnarray*}
and obtain the ($\kt$-dependent) BFKL splitting function as:
\begin{eqnarray*}
P_{BFKL} & = & \frac{6}{z} \Delta_{ns} = 6 z^{-1 + \omega}
\end{eqnarray*}
The BFKL evolution equation in integral form~\cite{Kwiecinski:1995pu,Kwiecinski:1996td} can be written as:
\begin{eqnarray}
\label{BFKL}
x{\cal A} (x,\kt,q )& =& x{\cal A}_0 (x,\kt,q) 
 +  \frac{\alphas}{2\pi} \int dz \frac{6}{z}\Delta_{ns}
  \int \frac{d q'^2}
{q'^{2}} 
\frac{x}{z}
{\cal A}\left(\frac{x}{z},\ktp,q'\right) \nonumber \\
& = & x{\cal A}_0 (x,\kt,q)  + \asb \int \frac{dz}{z} z^\omega  \int \frac{d q'^2}
{q'^{2}} 
\frac{x}{z}
{\cal A}\left(\frac{x}{z},\ktp,q'\right)
\end{eqnarray}
This equation describes parton emissions in the high energy limit, where all recoil effects can be neglected (leading order formulation), and thus is violating (in principle) energy-momentum conservation. Those effects are included as next-to-leading corrections. Recoil effects are also included in the formulation of the CCFM equation, which applies color coherence and includes the $1/(1-z)$ parts of the gluon splitting function, as formulated in Refs.~\cite{\ccfm,Marchesini:1990zy,Webber:1990rn}.  The CCFM evolution equation reads:
\begin{eqnarray}
x {{\cal A}}_a(x,{\bf k}, \mu^2) &= &\Delta_a (  \mu^2  ) \ 
 x{{\cal A}}_a(x,{\bf k},\mu^2_0)  \nonumber  \\
 & + &\int dz 
\int \frac{d^2 q}{\pi q^{2}} \Theta(\mu - zq) \frac{\Delta_a(\mu^2)}{\Delta_a( z^2 q^2) }
\tilde{P}(z,q,\kt) \frac{x}{z}{ {\cal A}}_b\left(\frac{x}{z}, {\bf k'} , {\bf q}^{\prime 2}\right) 
\label{CCFM_integral} 
\end{eqnarray}  
with $ {\bf k'} = {\bf k}+(1-z) {\bf q}^\prime$, and 
\begin{equation}
\tilde{P}(z,q,\kt)= \frac{\asb(q^2(1-z^2))}{1-z} + 
\frac{\asb(k^2_{t})}{z} \Delta_{ns}(z,q^2,k^2_{t})
\label{Pgg}
\end{equation}

The recently developed Parton Branching (PB) method \cite{Hautmann:2017fcj,Hautmann:2017xtx} is an application of eq.(\ref{TMD}) at NLO. Fits to  inclusive deep inelastic scattering data over a large range in $x$ and $Q^2$ leading to the first complete set of TMD distributions at NLO for quarks and gluons are reported in Ref.~\cite{Martinez:2018jxt}.
\begin{figure}[htbp] 
   \centering
   \includegraphics[width=6cm]{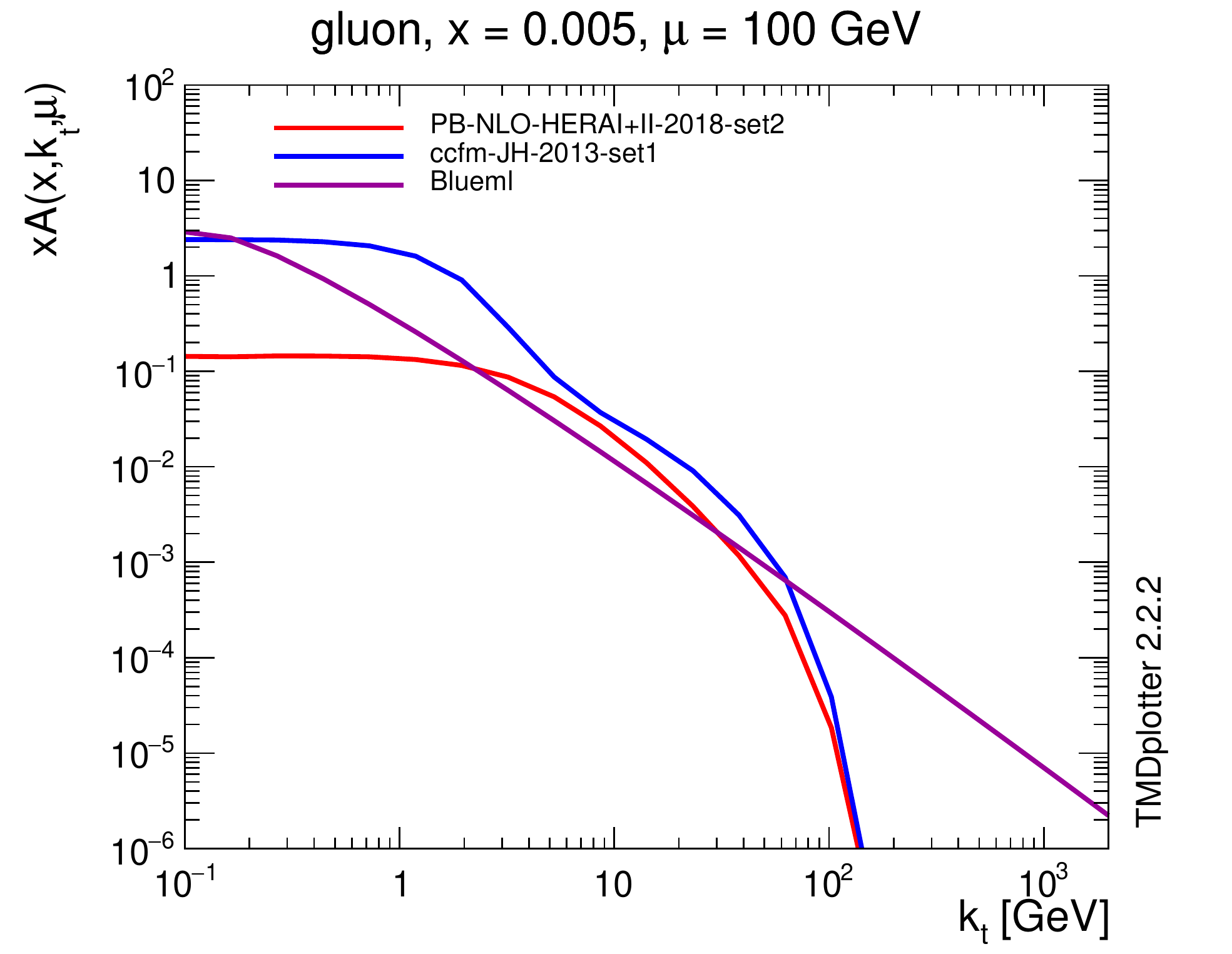}
   \includegraphics[width=6cm]{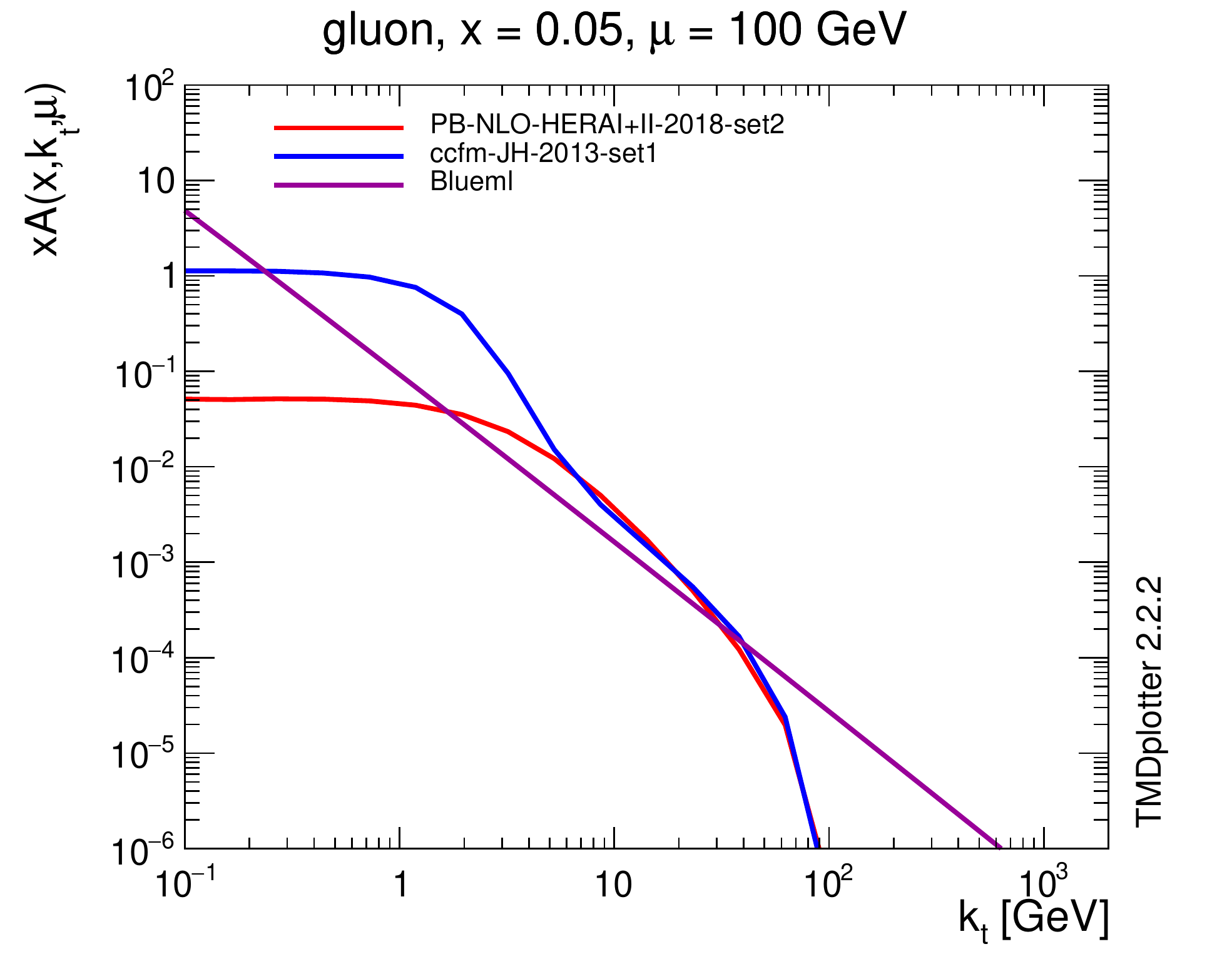}
   \caption{Unintegrated gluon densities obtained in the PB-method~\cite{Martinez:2018jxt}, CCFM~\cite{Hautmann:2013tba} and BKFL~\cite{Blumlein:1995mi} at different values of $x$ (obtained with TMDlib and TMDplotter~\cite{Abdulov:2021ivr,Connor:2016bmt,Hautmann:2014kza}).}
   \label{Fig:uGluon}
\end{figure}

The parton densities, and also the TMD parton densities are not observables themselves, only after convolution with the appropriate parton cross section, predictions can be made. Both BFKL and CCFM gluon densities need to be convoluted with off-shell matrix elements (as required from $\kt$-factorization~\cite{Catani:1993ww,Catani:1990eg,Catani:1990xk}), while the PB-TMD distributions are convoluted with on-shell ones. In addition, in BFKL and CCFM only gluon densities are calculated, in the PB-TMD calculation the full flavor set of quarks and gluons is determined at NLO. Keeping these differences in mind, it is still of interest to compare the $\kt$-dependent distributions for gluons (see Fig.~\ref{Fig:uGluon}). While both CCFM~\cite{Hautmann:2013tba} and PB-TMD~\cite{Martinez:2018jxt,Hautmann:2017fcj} distributions fall off quickly at $\kt \sim \mu$, the BFKL distribution~\cite{Blumlein:1995mi} has a tail to very large values of $\kt > \mu$.

\section{Measurements with sensitivity to BFKL effects}
\begin{figure}[htbp] 
   \centering
   \includegraphics[width=3cm]{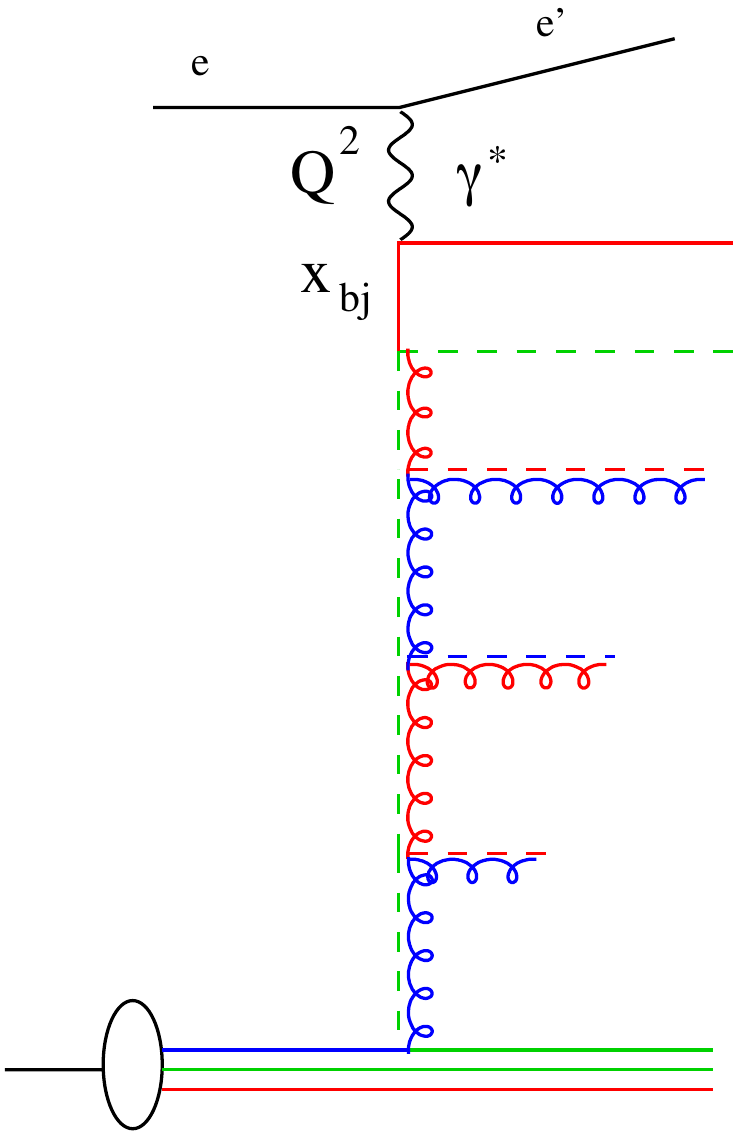}
   \includegraphics[width=3cm]{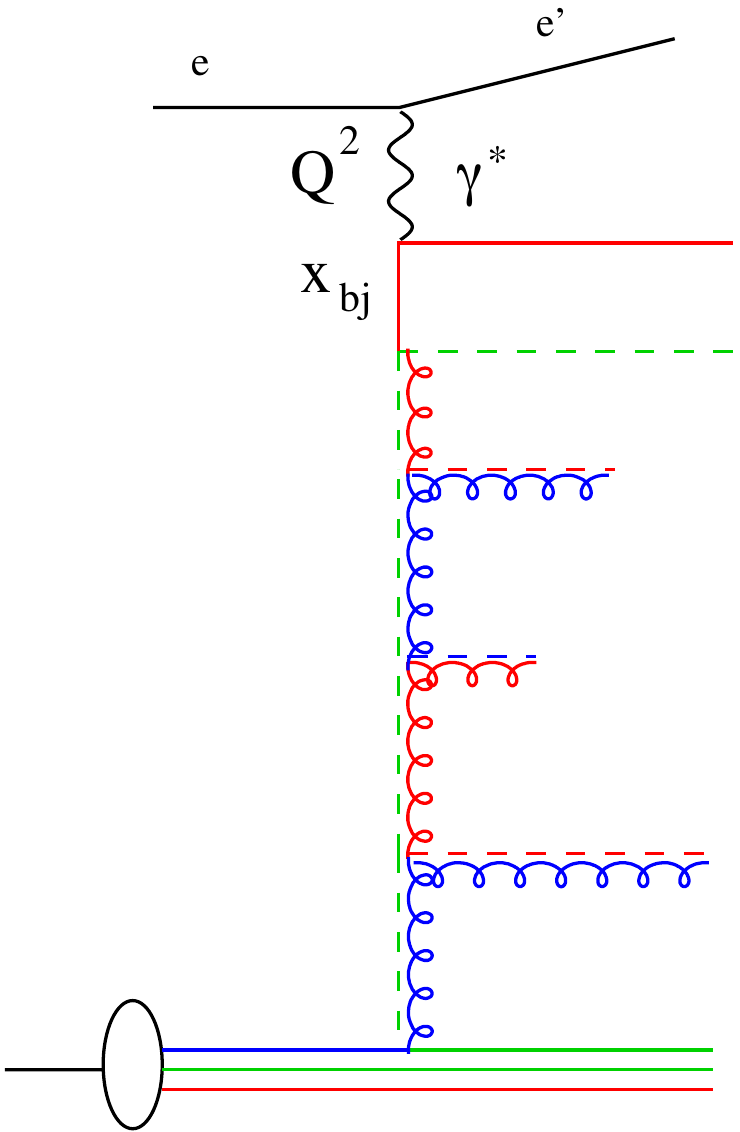}
   \caption{Schematic illustration of a DGLAP-evolution (left) and a BFKL evolution (right). The length of the gluons indicates the size of their transverse momenta.}
   \label{Fig:DGLAP_BFKL}
\end{figure}

While parton emissions from the initial state described by DGLAP evolution (i.e. its formulation in terms of exclusive state, like parton showers or with the PB method) are essentially ordered in transverse momentum of the emitted partons (or in its variant with angular ordering), emissions described by BFKL or CCFM evolution at small $x$ can perform a random walk in transverse momentum (only limited by overall kinematic constraints), shown schematically in Fig.~\ref{Fig:DGLAP_BFKL}.
Partonic emissions in a BFKL evolution will have a different transverse momentum spectrum at a fixed rapidity, compared to those coming from a DGLAP evolution. Differences are expected in particular close to the proton direction, where the transverse momentum coming from a DGLAP evolution is expected to be small.

In the following sections examples of measurements performed at the $ep$ collider HERA at DESY are discussed, which probe parton kinematics at small values of~$x$.

\subsection{Inclusive cross section measurements}\label{sec:inclusive_xsection}
\begin{figure}[htbp] 
   \centering
   \includegraphics[width=5.cm]{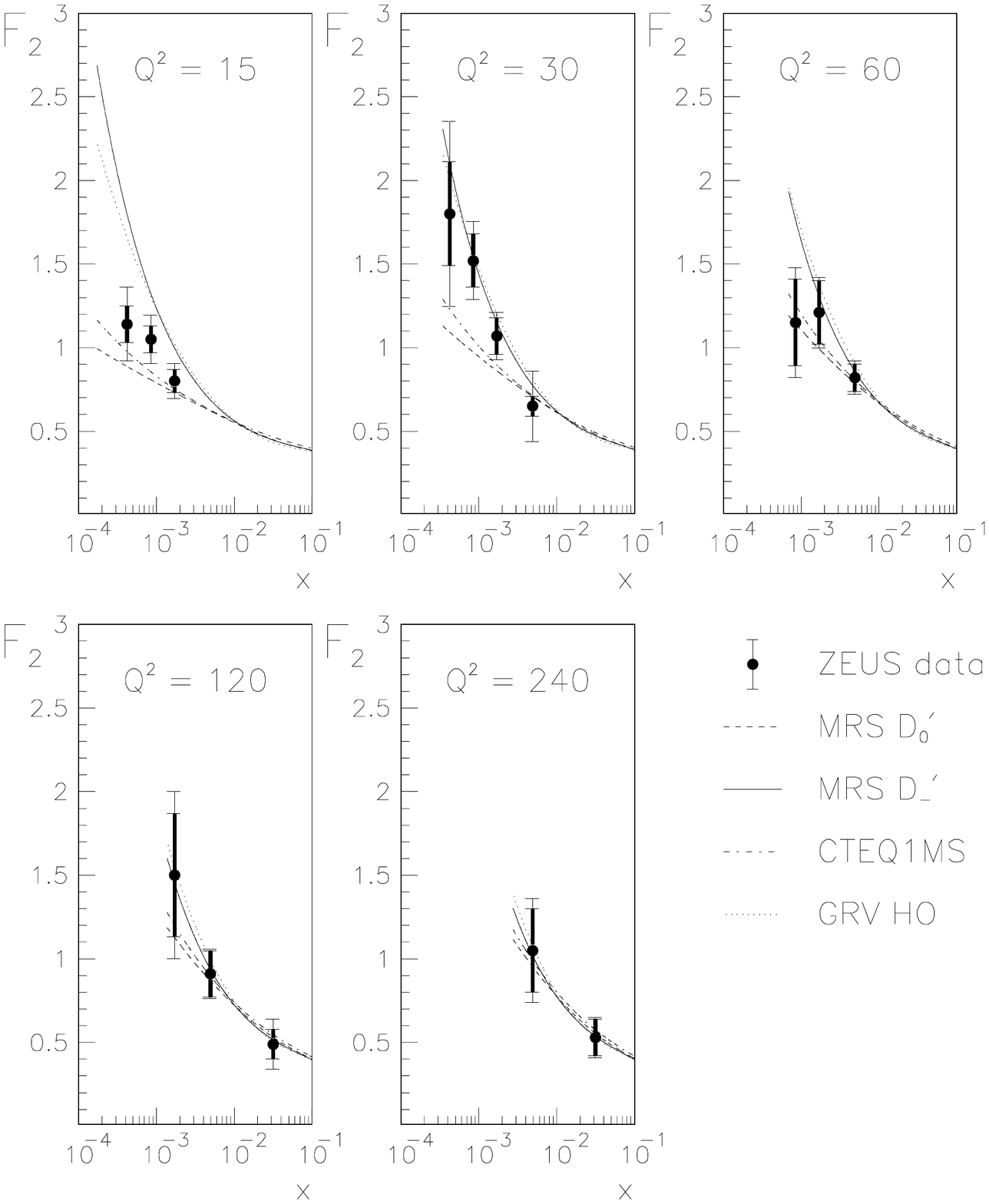} 
    \includegraphics[width=7.cm]{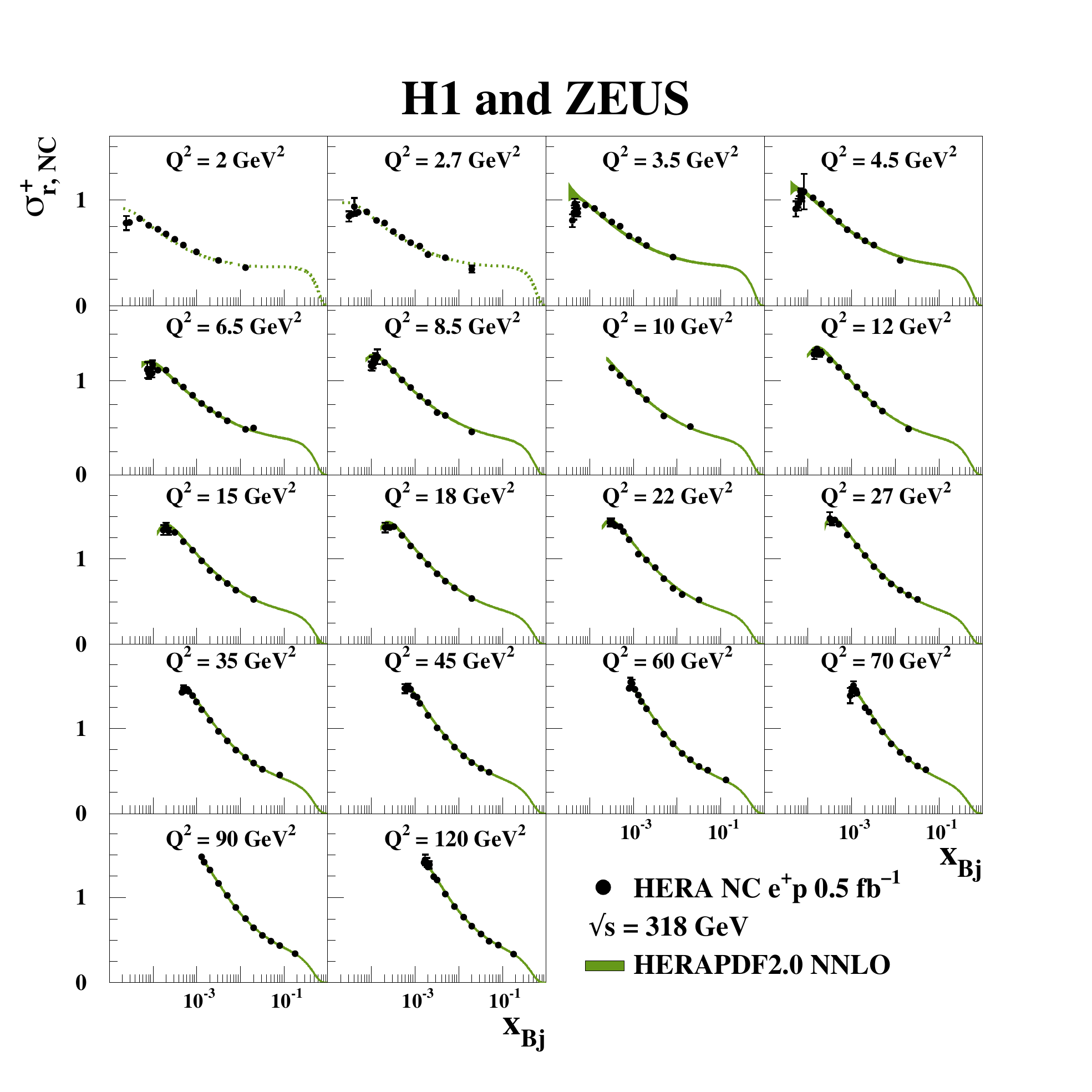}
   \caption{Measurements of deep inelastic scattering cross section: (left) first measurement by of the proton structure function $F_2(x,Q^2)$~\protect\cite{Derrick:1993fta}, (right) final measurement of the reduced cross section obtained from a combination of the measurements from H1 and ZEUS~\protect\cite{Abramowicz:2015mha} }
   \label{Fig:F2}
\end{figure}

The inclusive cross section for deep-inelastic scattering was measured already at the start of HERA in 1992 \cite{Abt:1993cb,Derrick:1993fta},
an example of this very first measurement of $F_2(x,Q^2)$ from ZEUS is shown  in Fig.~\ref{Fig:F2} (left). The behavior of $F_2(x,Q^2)$ at small values of the momentum fraction $x$ indicates a clear rise, in contrast to some of the predictions.

In Fig.~\ref{Fig:F2} (right) the reduced cross section ($\sigma_{red}(x,Q^2) \sim F_2(x,Q^2)$)  from a combination of the measurements of H1 and ZEUS after the full HERA  running \cite{Abramowicz:2015mha} is shown, confirming impressively the rise of the cross section for small values of $x$, but also showing the enormous improvement in precision of the measurements from the start of HERA to the finally combined results at the end of the HERA running.
The rise of $F_2$ at small $x$ was seen as an indication for BFKL effects (for a discussion see \cite{Bartels:1992jf,Bartels:1990zk,Bartels:1989rj,Kwiecinski:1997vc,Kwiecinski:1997ee,Kwiecinski:1995pw,Kwiecinski:1995pu,Kwiecinski:1991cb}), however,
the measurements of the inclusive deep inelastic cross section ($\sigma_{red}$) could be well described applying the DGLAP evolution at next-to-leading (NLO) as well as at next-to-next-to-leading order in the expansion of the strong coupling $\alpha_s$, as shown in Fig.~\ref{Fig:F2} (right). The fits performed in Ref.~\cite{Abramowicz:2015mha} show a very good $\chi^2/ndf$ when measurements with $Q^2 > 3.5 $~GeV$^2$ are included.
\begin{figure}[htbp] 
   \centering
   \includegraphics[width=9cm]{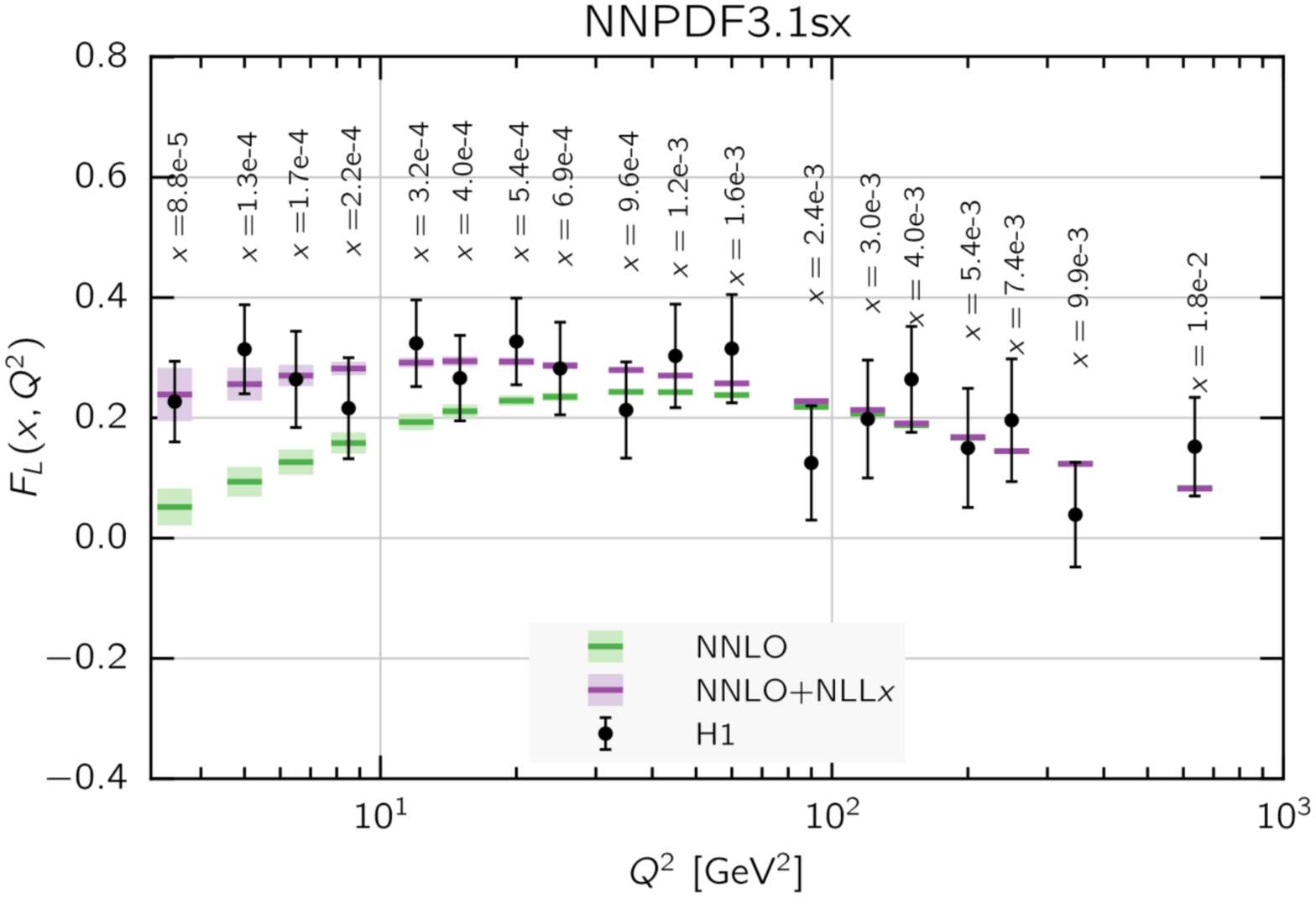}
   \caption{Measurement of the longitudinal structure function $F_L(x,Q^2)$ and comparison with predictions using small $x$ resummation~\protect\cite{Ball:2017otu}.}
   \label{Fig:FLsmallx}
\end{figure}

Recently a dedicated study, including BFKL resummation to the description of the inclusive cross section, proved the need for BFKL effects \cite{Ball:2017otu,Abdolmaleki:2018jln}. The effect of small $x$ resummation is shown for the longitudinal structure function $F_L(x,Q^2)$ in Fig.~\ref{Fig:FLsmallx} (taken from \cite{Ball:2017otu}).
This study showed for the first time, that the measurements can be described  better, if BFKL effect are included in the calculation.

\subsection{Transverse energy flow}
\begin{figure}[htbp] 
   \centering
   \includegraphics[width=10cm]{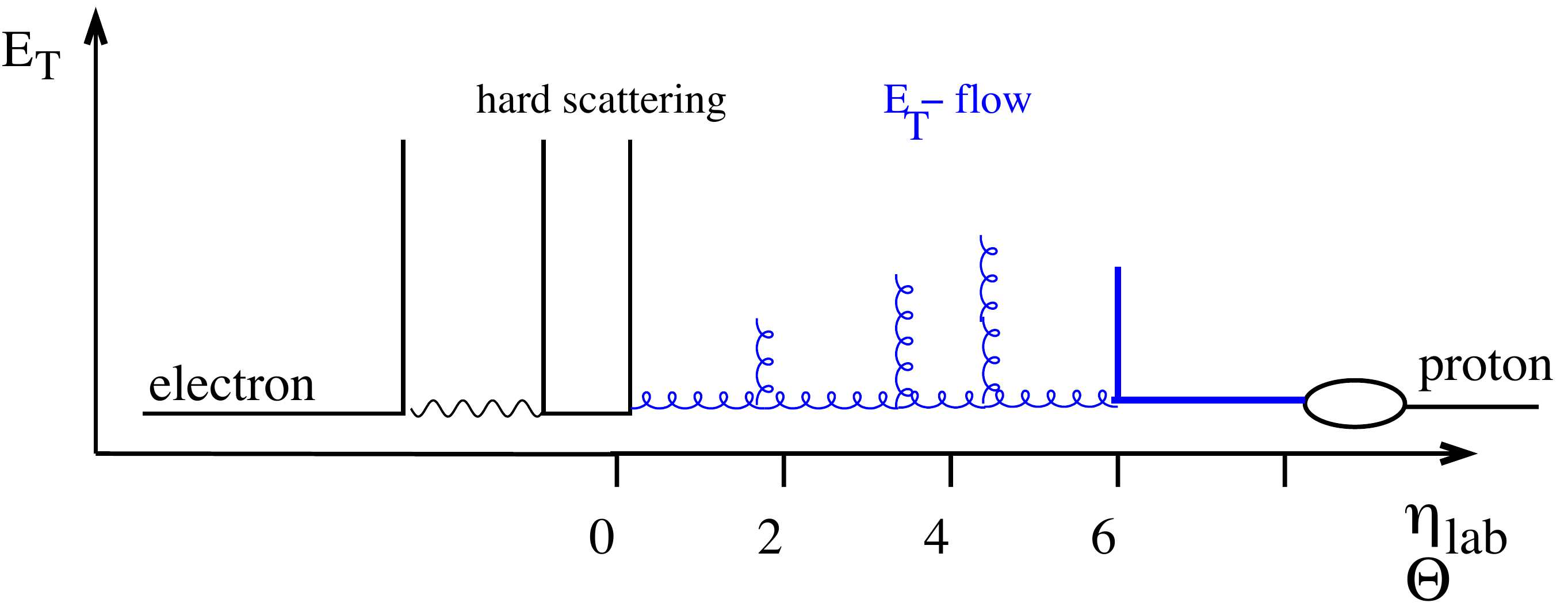}
   \caption{Schematic view of the process contributing to transverse energy flow in deep inelastic scattering.  }
   \label{Fig:Et_flow-schematic}
\end{figure}

A measurement of the transverse energy flow as a function of the rapidity can give insights in the way transverse momenta of  partons along the evolution chain are distributed (as illustrated in Fig.~\ref{Fig:Et_flow-schematic}). In a DGLAP scenario, the transverse momenta of partons close to the proton direction are expected to be small and they increase towards the hard scattering, while in a BFKL scenario the transverse momenta can be sizable all over the rapidity range, and can form a so-called {\it Bartels-cigar}\cite{Bartels:1995yk}, a diffusion pattern of the transverse momenta.

\begin{figure}[htbp] 
   \centering
   \includegraphics[width=5cm]{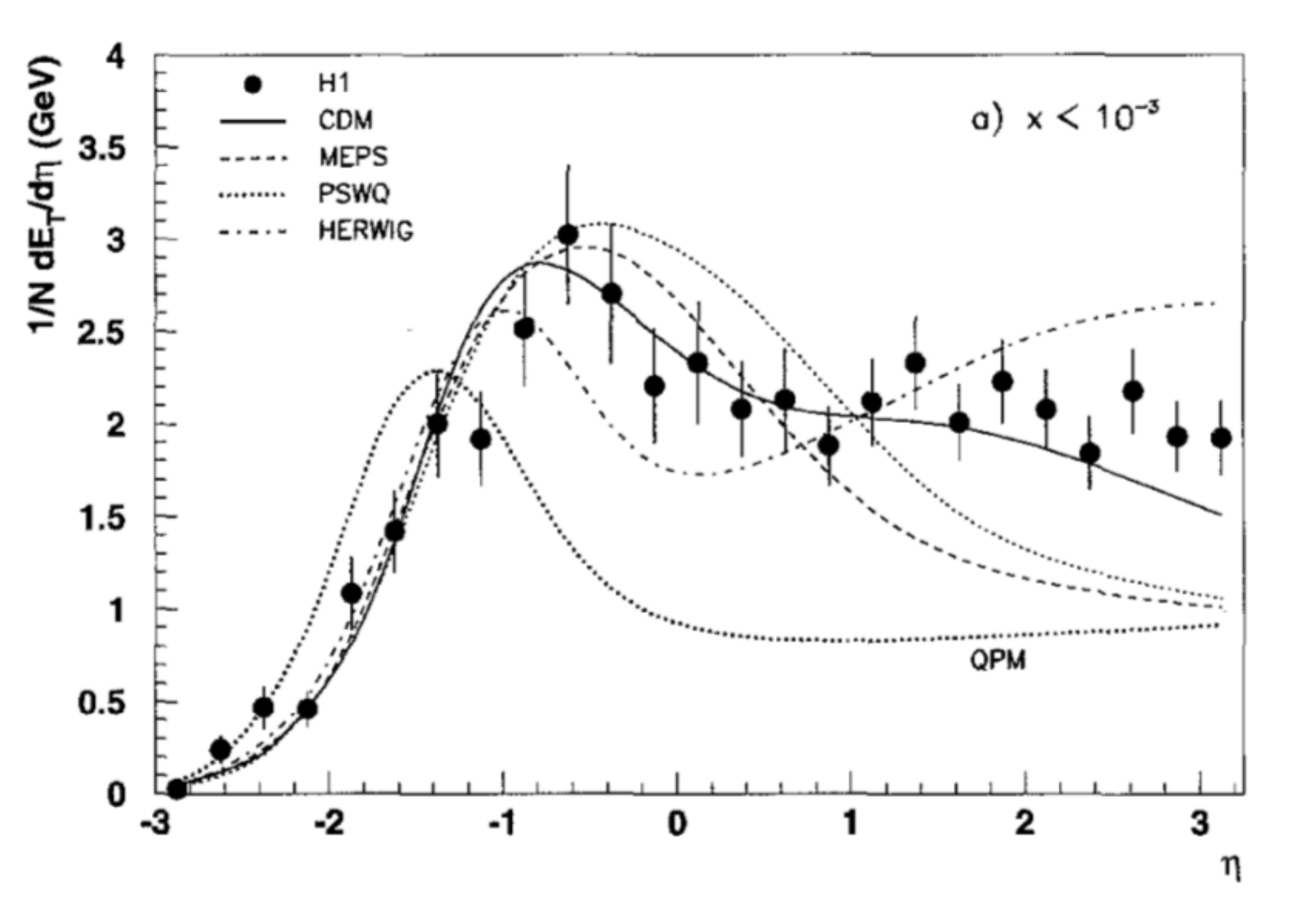}
   \includegraphics[width=7.3cm]{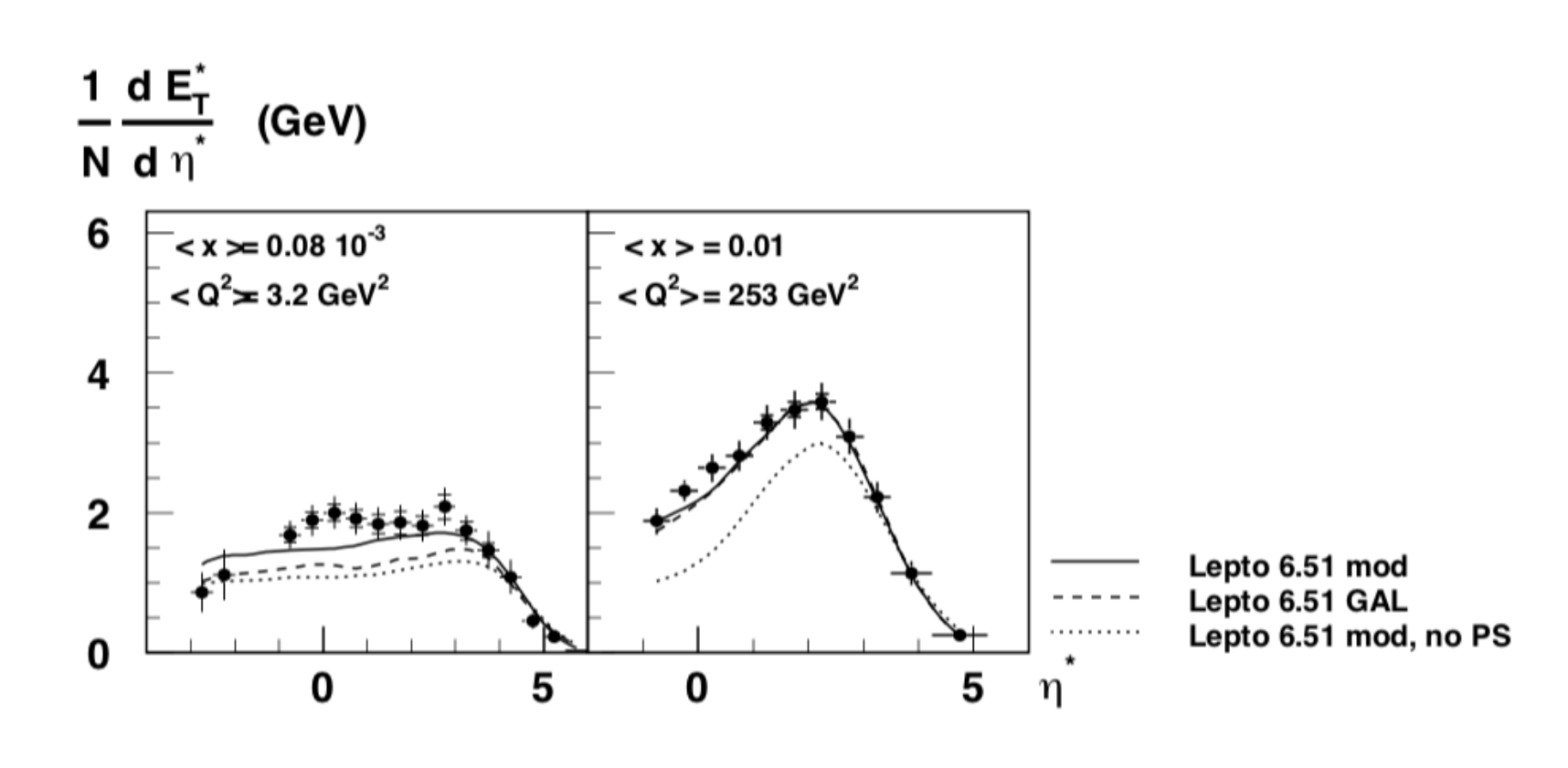}
   \caption{Transverse energy flow in deep inelastic scattering: (left) the distribution as a function of the difference to the direction of the scattered quark is shown (left) for the first measurement~\cite{Abt:1994ye}, (right) the distribution with higher statistic, and for different regions of $x$, is shown ~\cite{Adloff:1999ws}.  }
   \label{Fig:Et_flow}
\end{figure}

The transverse energy flow in deep inelastic scattering was measured by H1\cite{Abt:1994ye} and ZEUS\cite{Derrick:1993gz}. A significant transverse energy was found away from the hard scattering.  An example of one of the first measurements is shown in Fig.~\ref{Fig:Et_flow} (right), and compared to predictions based on simulations using parton showers.
All predictions\footnote{The prediction from HERWIG~\cite{Marchesini:1991ch} had issues with large cluster sizes.} fall below the measurements at large rapidities, except the prediction labelled as CDM~\cite{Lonnblad:1992tz} (which simulates unordered emissions, similar to what is expected from BFKL). These first measurements were also compared to calculations (i.e. in Ref.~\cite{GolecBiernat:1994fw}) obtained with BFKL evolution, which confirmed the increased energy flow in the forward region.
Later measurements~\cite{Adloff:1999ws} with much higher statistics (shown in Fig.~\ref{Fig:Et_flow} (right)) and with a finer binning in $x$ and $Q^2$ confirmed these early findings (note that here the energy flow is plotted in the hadronic center-of-mass system, and the proton direction appears at small $\eta^*$).
All the different variants of the simulation program LEPTO~\cite{Ingelman:1996mq} (which includes order $\as$ matrix elements and parton showers) fall below the measurement at $\eta^* \sim 0$ at small $x$ and small $Q^2$.

\subsection{Forward jet and forward $\boldmath\pi^0$ cross section}
\begin{figure}[htbp] 
   \centering
   \includegraphics[width=5cm]{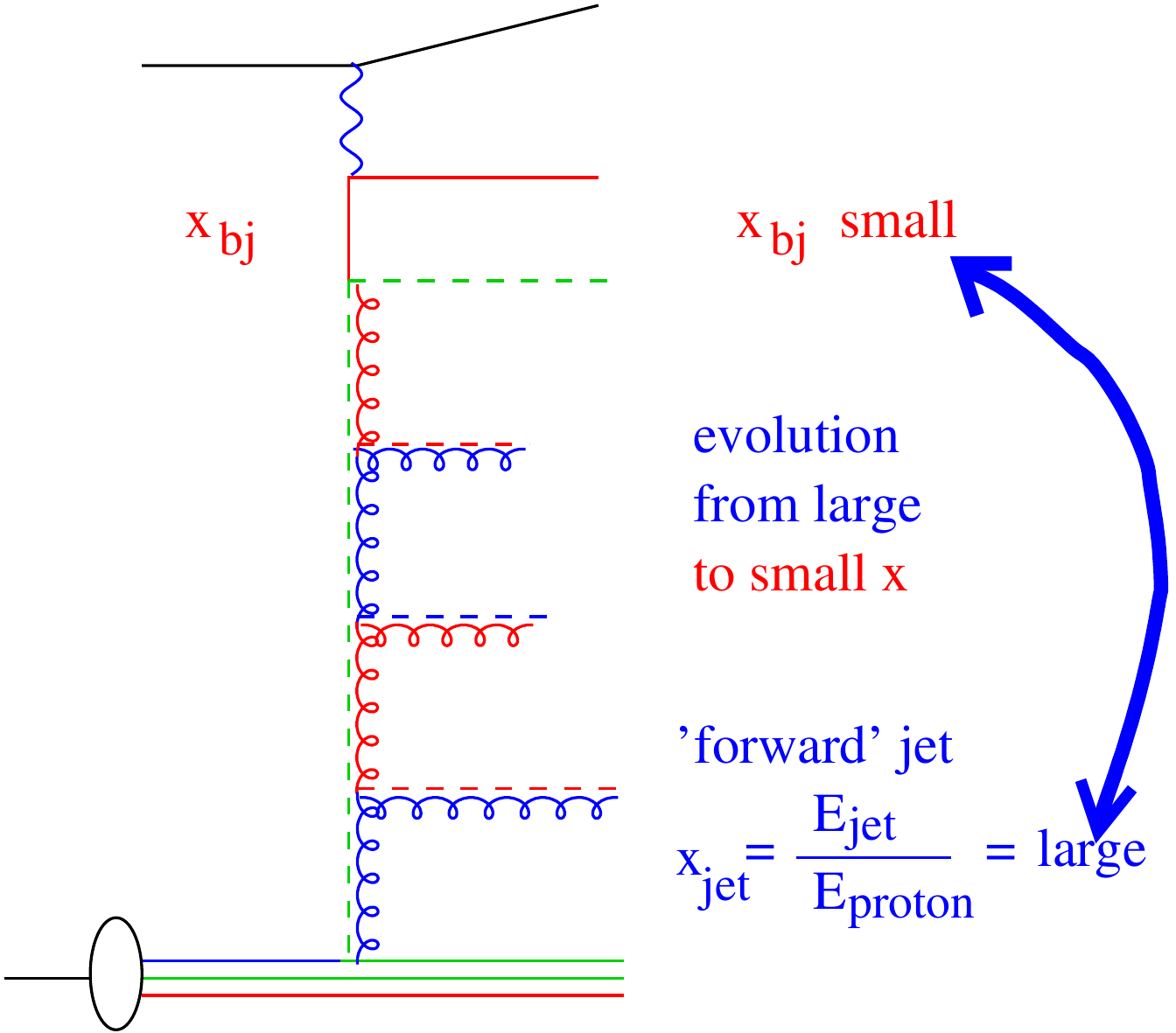}
   \includegraphics[width=6cm]{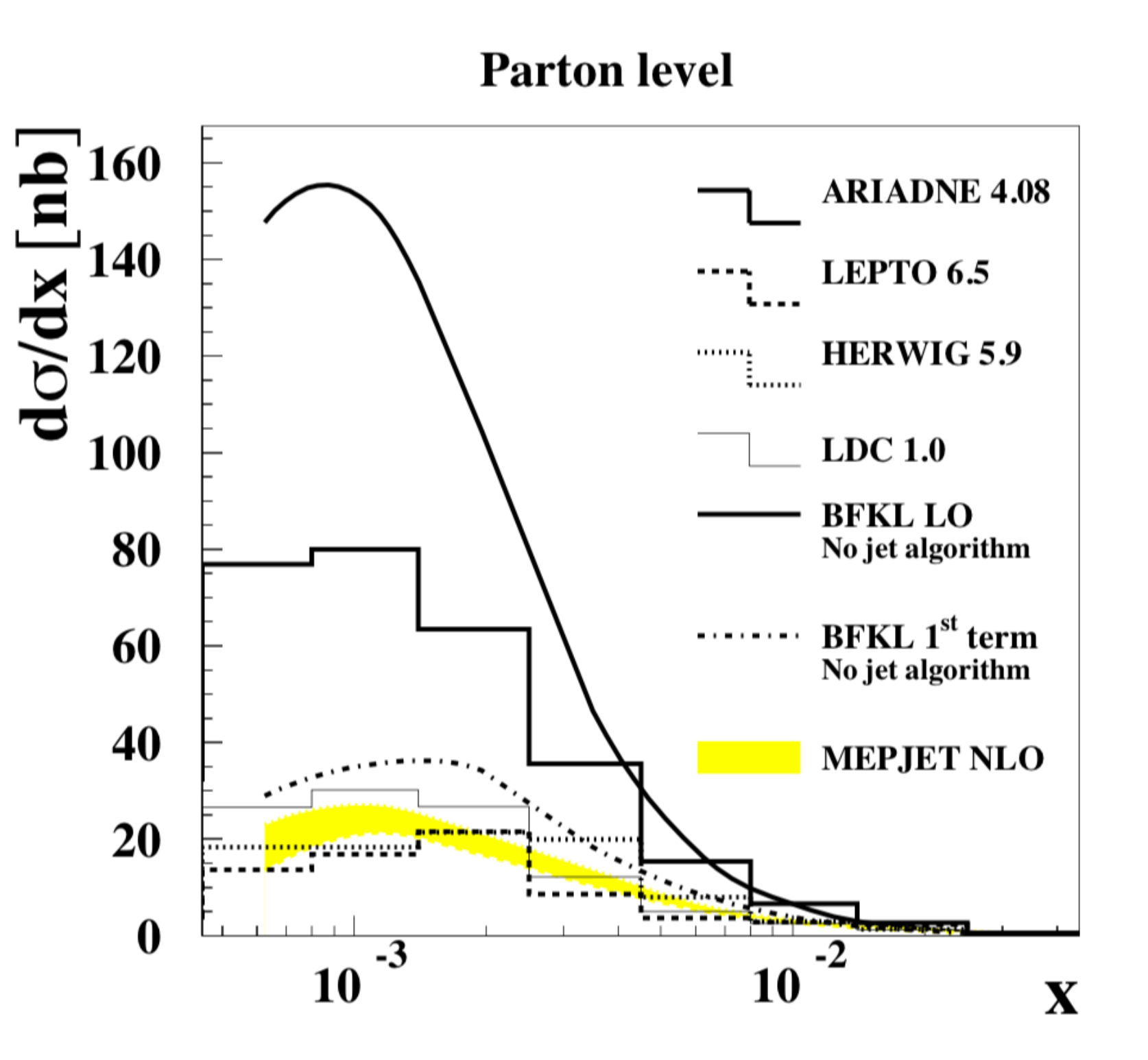}
   \caption{Left: schematic diagram illustrating forward jet production. Right: predictions for forward jet production calculated in different schemes~\cite{Breitweg:1998ed}}
   \label{Fig:fwdjets}
\end{figure}

Measurements dedicated to investigate BFKL dynamics were proposed in Refs.~\cite{Mueller:1990er,Mueller:1990gy}. The idea is to measure the cross section as a function of $x$ for a jet in the forward region (at large $x_{jet}$) with a transverse momentum similar to the virtuality of the photon. By this requirement, DGLAP evolution is expected to be suppressed, while the $\kt$-unordered evolution from BFKL will give significant contributions, at least at small $x$. The diagram for forward jet production is schematically shown in Fig.~\ref{Fig:fwdjets} (left). In  Fig.~\ref{Fig:fwdjets} (right) calculations for this process are shown: except the predictions labelled as {\it BFKL LO}~\cite{Bartels:1996wx,Bartels:1996gr} and ARIADNE (which is the same calculation mentioned in the previous section as CDM) all predictions give a rather small cross section for forward jet production. 
The forward jet cross section was also of interest for investigations of {\it hot-spots}, localized regions with high density, inside the proton \cite{Bartels:1991tf}.

\begin{figure}[htbp] 
   \centering
   \includegraphics[width=6.5cm]{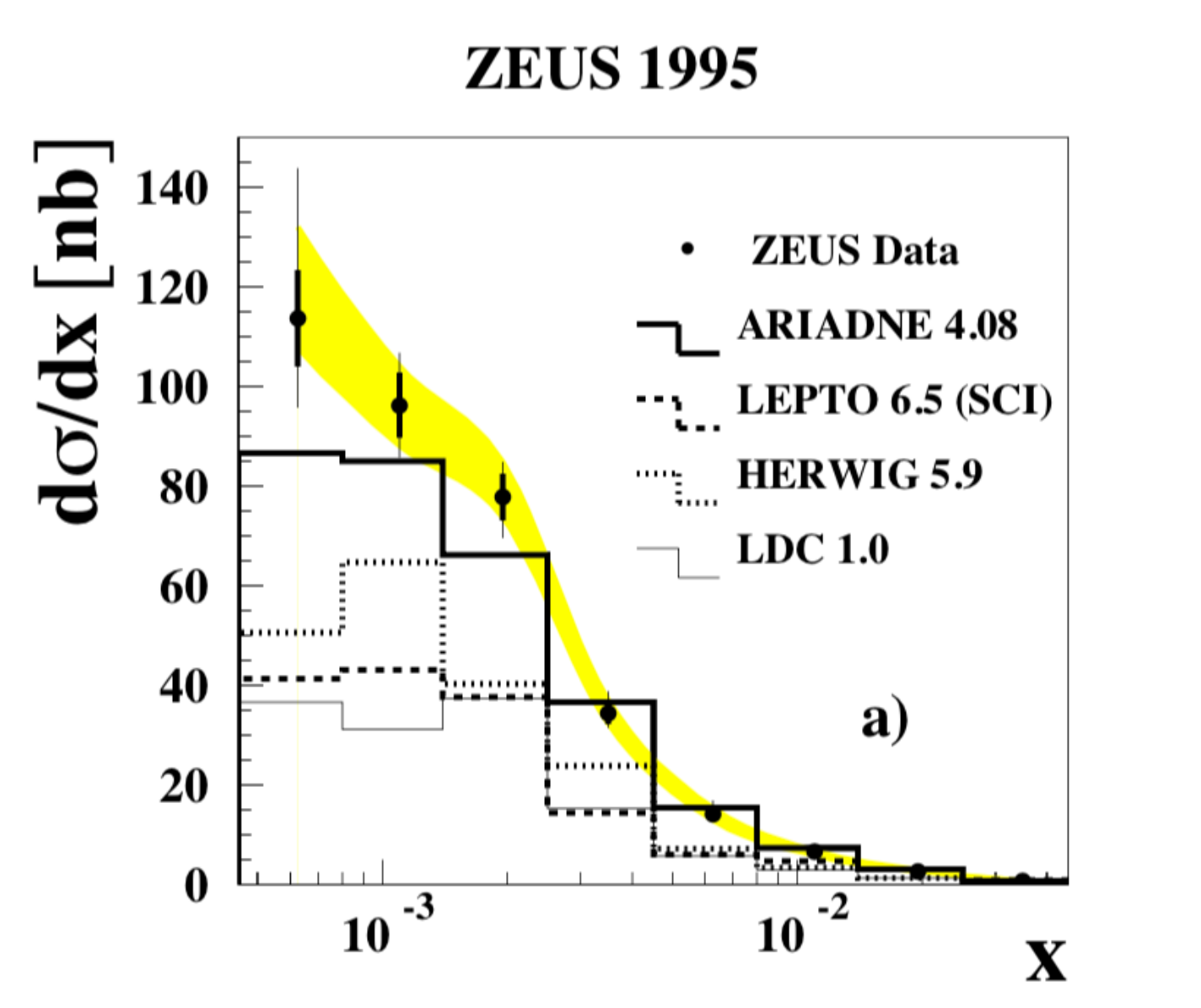}
   \includegraphics[width=6cm]{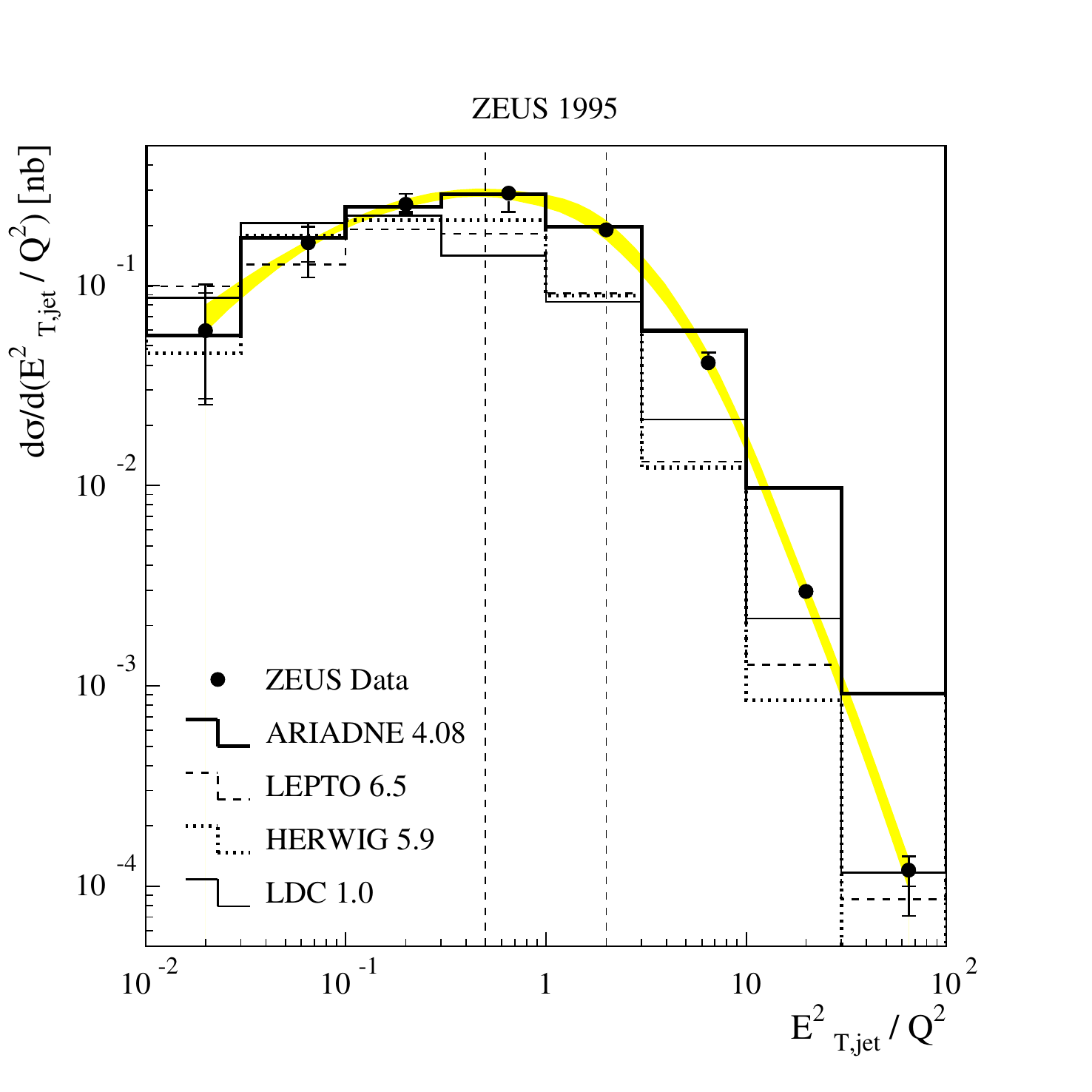}
   \caption{Left: measurement of forward jet production as a function of $x$~\cite{Breitweg:1998ed}.  Right: jet production cross section as a function of $E_T^2/Q^2$~\cite{Breitweg:1999ss}.}
   \label{Fig:fwdjetsZEUS}
\end{figure}

Forward jet production was measured by H1~\cite{Aid:1995we,Adloff:1998fa,Aktas:2005up,Aaron:2011ef} and ZEUS~\cite{Breitweg:1998ed,Breitweg:1999ss,Chekanov:2005yb,Chekanov:2007pa} collaborations. Measurements from ZEUS are shown in Fig.~\ref{Fig:fwdjetsZEUS} for jets with $p_T > 5 $~GeV. In  Fig.~\ref{Fig:fwdjetsZEUS} (left) it is clearly seen, that the measured cross section increases steeply towards small $x$, while the predictions obtained from DGLAP parton shower simulations fall below the measurements. Only the prediction from ARIADNE comes closer the the data.
In Fig.~\ref{Fig:fwdjetsZEUS} (right) the cross section is shown as a function of $E_T^2/Q^2$, with $E_T$ being the transverse momentum of the forward jet. For the typical DGLAP region, where $E_T^2$ is small compared to $Q^2$, the predictions describe the measurement, while in the region where $E_T^2$ becomes closer or even larger than $Q^2$, the measurements are above the predictions, indicating a very  different kinematic region. 

\begin{figure}[htbp] 
   \centering
   \includegraphics[width=10cm]{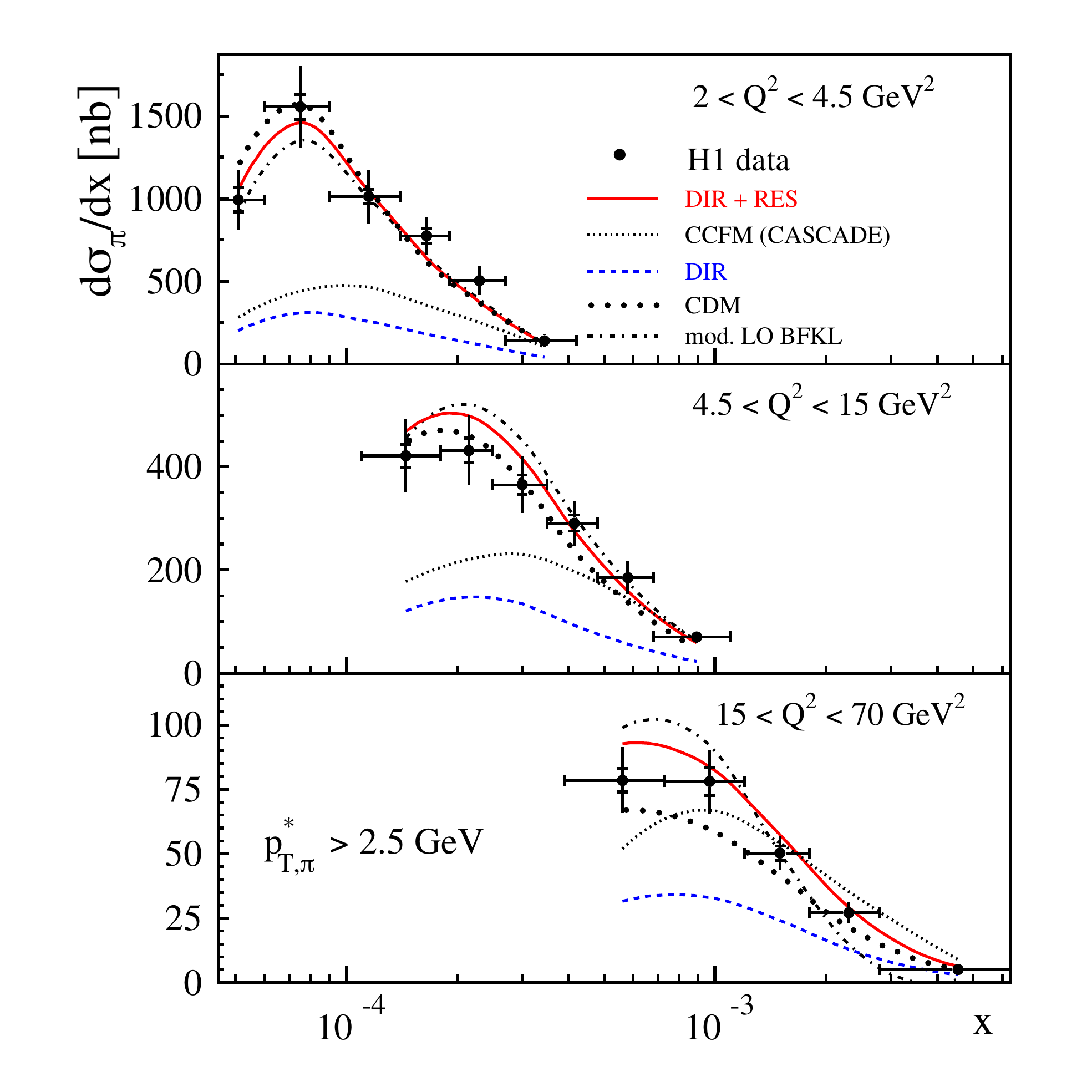}
   \caption{Measurement of forward $\pi^0$ production as a function of $x$ and different $Q^2$~\cite{Aktas:2004rb}.}
   \label{Fig:fwdpion}
\end{figure}

One of the experimental issues in forward jet production is the requirement of the transverse momentum of the jet, i.e. $p_T > 5$~GeV, which limits the kinematic range, but is necessary for a proper jet identification. Identified particles like $\pi^0 \to \gamma \gamma$ have been used in H1 to measure a similar process~\cite{Adloff:1999zx,Aktas:2004rb}, but now being able to access lower $x$ and lower transverse momenta. The measurements from H1 are shown in Fig.~\ref{Fig:fwdpion} and compared to predictions using RAPGAP~\cite{Jung:1993gf,RAPGAP32} (which includes order $\as$ matrix elements, and in addition contributions from resolved virtual photons~\cite{Jung:1998fu,Jung:1998wh}). The model including resolved virtual photons describes the measurement very well, while a prediction based on the CCFM evolution equation \cite{Jung:2001hx,Jung:2000hk} falls below the measurements.

The forward jet and forward pion cross sections increase towards small $x$, and only predictions which allow for unordered emissions during the initial state cascade come close to the measurements, while predictions based on DGLAP parton showers, even with order $\as$ matrix elements are significantly below the data. {\it Hot-spots} inside the proton could not be identified within the experimental precision reached so far.

\subsection{Charged Particle Spectra}
\begin{figure}[htbp] 
   \centering
   \includegraphics[width=4cm]{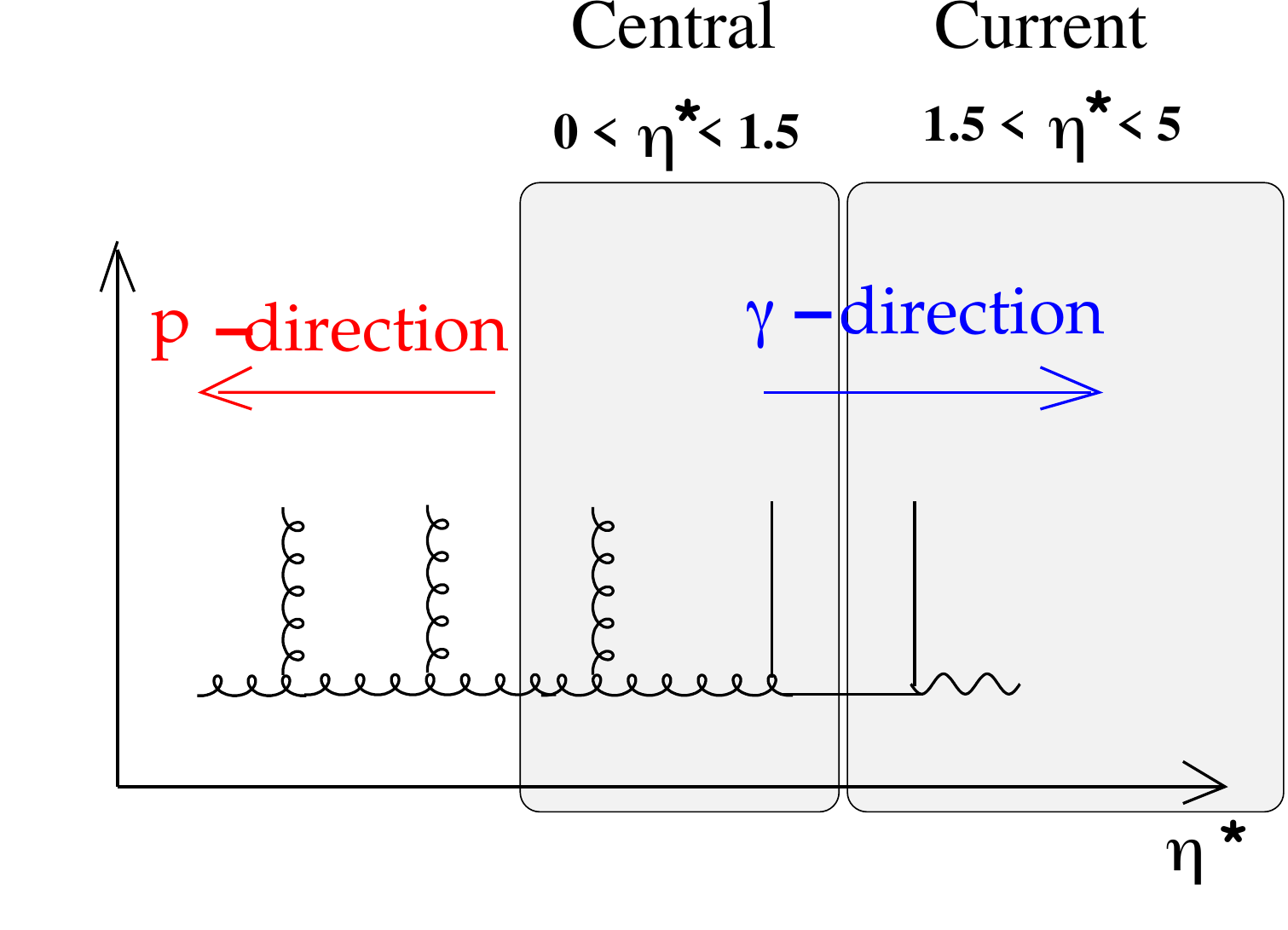}
   \caption{Schematic illustration of the phase space region used for  the measurement of charged particle spectra~\cite{Alexa:2013vkv}}
   \label{Fig:charged_particlesSchematic}
\end{figure}

\begin{figure}[htbp] 
   \centering
   \includegraphics[width=6cm]{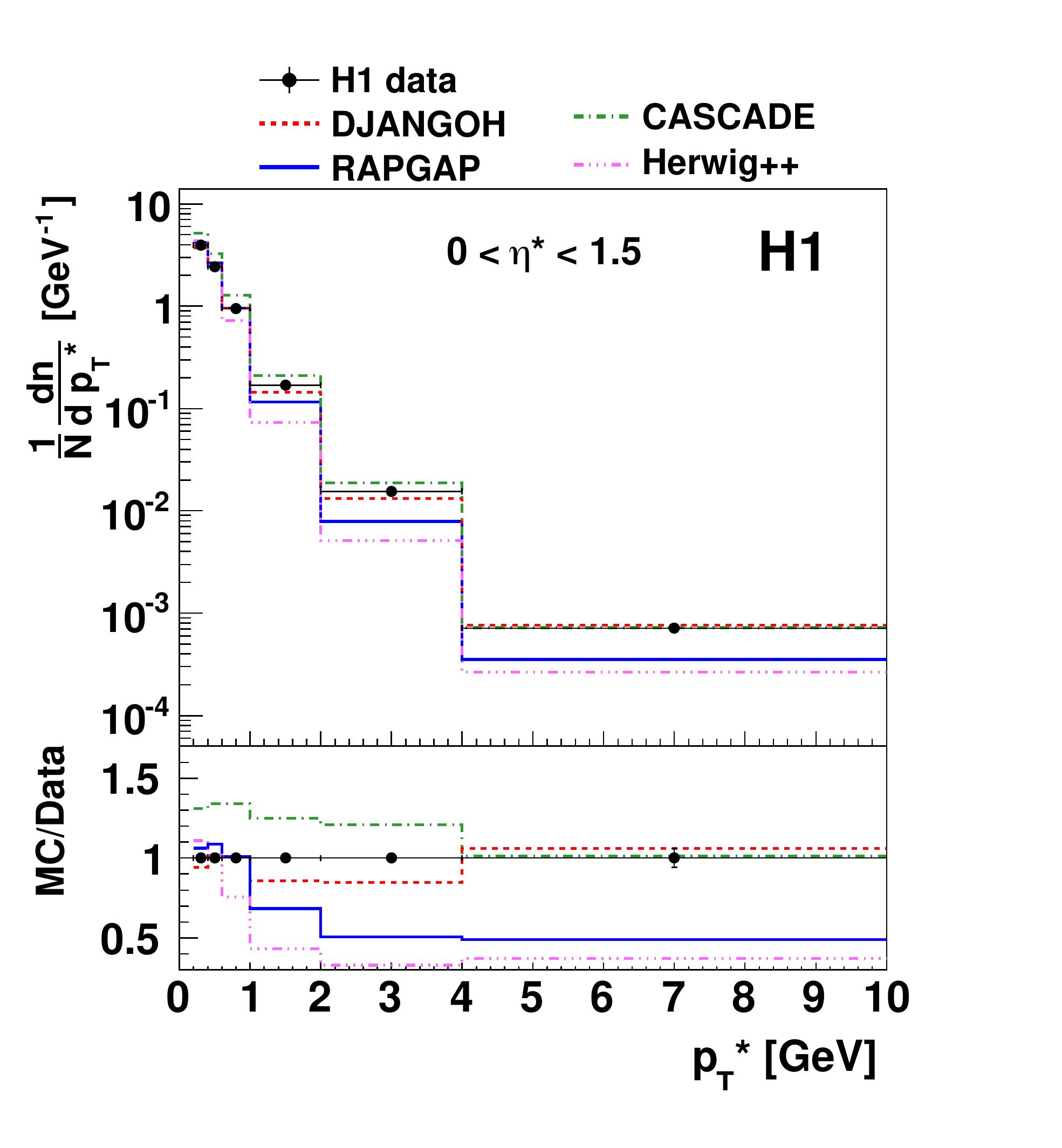}
   \includegraphics[width=6cm]{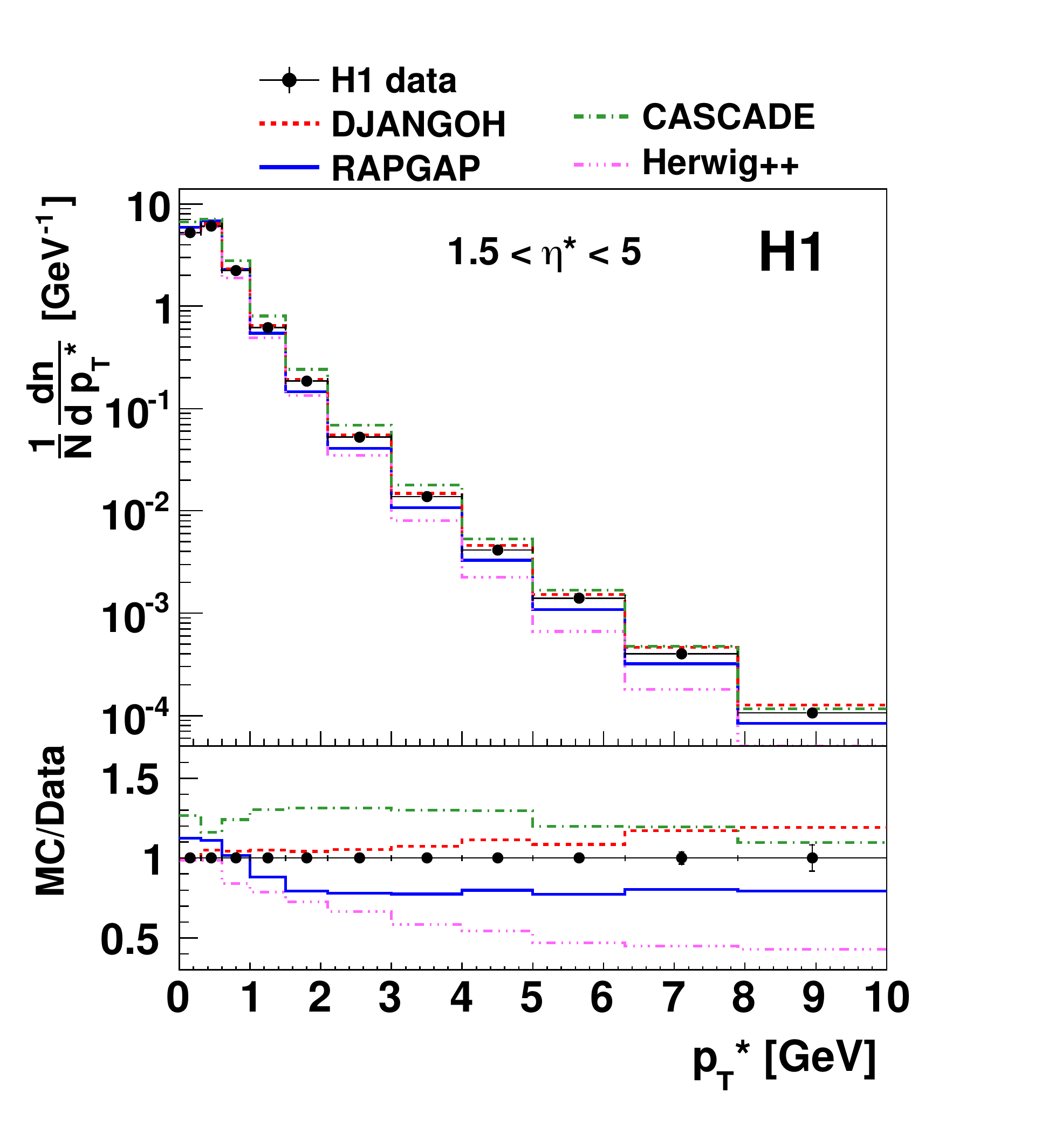}
   \caption{Transverse momentum spectrum of charged particles~\cite{Alexa:2013vkv}. Left: the region where the parton evolution dominates (proton direction). Right: the region where the hard process dominates (photon direction)}
   \label{Fig:charged_particles}
\end{figure}

A new type of measurements, based on inclusive transverse momentum spectra of charged particles was proposed in Ref.~\cite{Kuhlen:1996et}. The advantage of such measurements is that one is able to probe  very small transverse momenta, accessing very low $x$ values. The measurement is performed in the hadronic center of mass frame, as indicated in Fig.~\ref{Fig:charged_particlesSchematic}. The measurement is shown in Fig.~\ref{Fig:charged_particles} (left) for the proton direction and Fig.~\ref{Fig:charged_particles} (right) for the photon direction region. In the proton direction region, the predictions based on DGLAP based parton shower simulations (DJANGOH~\cite{Schuler:1991yg}, RAPGAP~\cite{Jung:1993gf,RAPGAP32} , HERWIG++~\cite{Bahr:2008pv}) are significantly below the measurements for transverse momenta $p_T > 1 $~GeV, while in the photon direction region the predictions (except HERWIG++) come closer to the measurements. The calculation based on CCFM (CASCADE~\cite{Jung:2010si,Jung:2001hx}) comes closer to the measurement in the proton direction region at larger transverse momenta.

\subsection{Fixed order calculations and small \boldmath$x$ effects}
Most of the calculations presented so far were based on order $\as$ matrix elements supplemented with parton showers. Calculations using the complete next-to-leading order matrix elements, matched with parton showers, became available only very recently \cite{Bellm:2015jjp}. However, with advanced techniques of matching and merging of higher order tree-level diagrams, calculations were performed for  deep inelastic scattering~\cite{Carli:2010cg} using tree-level diagrams of up to 5 partons. 

\begin{figure}[htbp] 
   \centering
   \includegraphics[width=10cm]{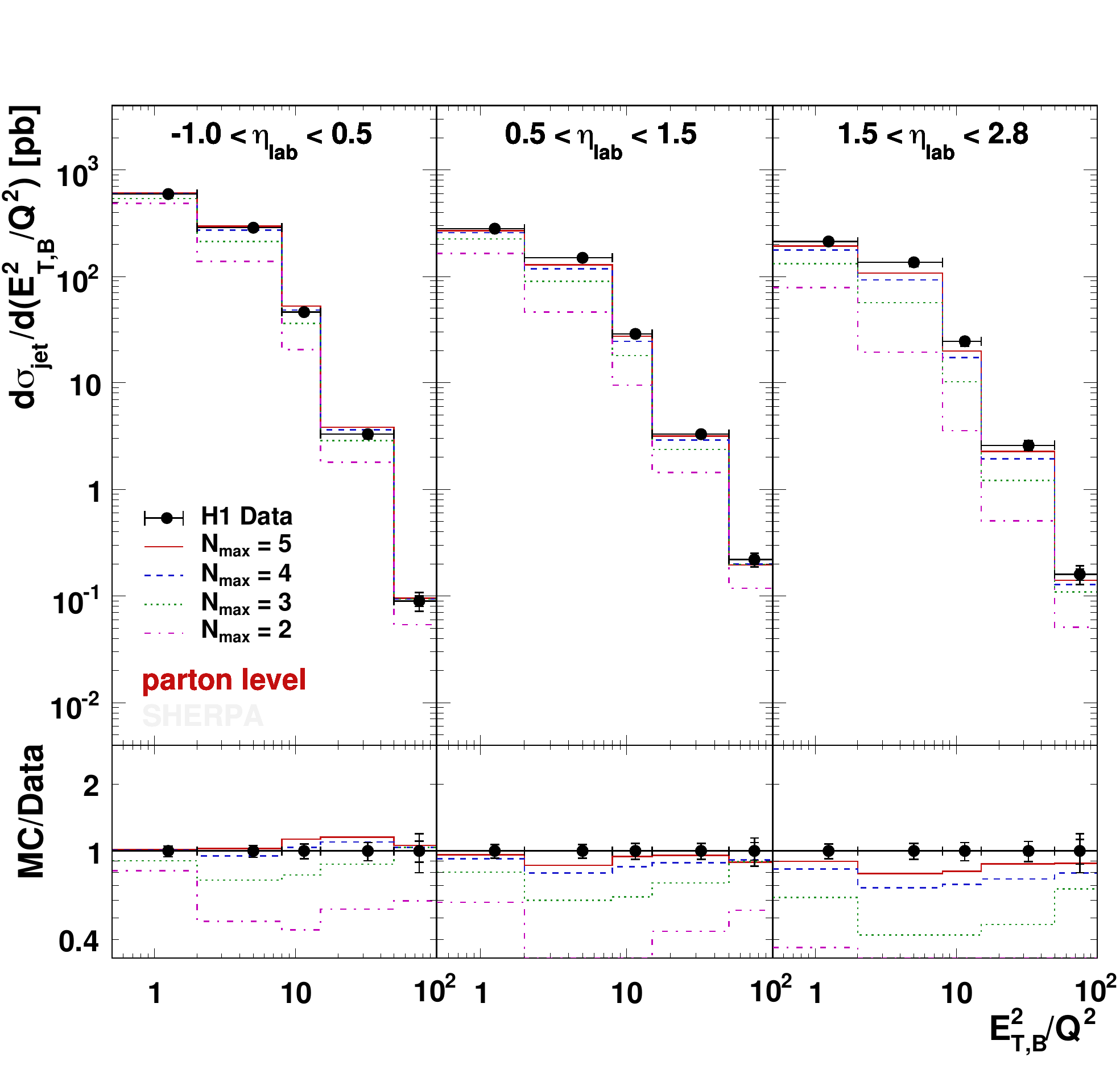}
   \caption{Measurement of inclusive jet production from H1~\cite{Adloff:2002ew} compared with predictions based on different parton multiplicities of the matrix element calculations~\cite{Carli:2010cg}.}
   \label{Fig:MultiJetMerged}
\end{figure}

In Fig.~\ref{Fig:MultiJetMerged} a measurement~\cite{Adloff:2002ew} of inclusive jet production as a function of $E_T^2/Q^2$ is shown for different rapidity regions of the jet, compared to predictions based on different parton multiplicities of the matrix elements.
It is interesting to observe, that (at least) some measurements dedicated to observe BFKL effects are  well described when higher order matrix elements are included in the calculations. Matrix element calculations cover the full available phase space (since no approximations are applied), but it remains to be seen, whether a fixed order calculation is sufficient to describe the measurements at higher energies, or whether resummation of enhanced contributions to all orders, as performed with BFKL or CCFM calculations, are required.

\section{The need for full hadronic Monte Carlo event generators for small \boldmath$x$ physics.}

Apart from inclusive cross section measurements (as described in section~\ref{sec:inclusive_xsection}), comparisons with measurements require the simulation of complicated cuts in phase space, and event generators are required. Some event generators provide only the partonic configuration \cite{Nagy:2001xb,Currie:2017tpe}, but for many applications a simulation of the complete final state is needed. Most of the event generators applied at HERA were using collinear parton densities and matrix elements up to order $\as$ to calculate the cross sections supplemented with parton showers based on the DGLAP evolution equation and hadronization (LEPTO, HERWIG and RAPGAP). The package ARIADNE applied color dipoles to simulate the partonic cascade. Only the generators LDCMC~\cite{Andersson:1995jv,Gustafson:2002kz} and CASCADE~\cite{Jung:2010si,Jung:2001hx} used directly a BFKL motivated parton shower, in the form of the CCFM formulation. The CASCADE event generator applied unintegrated parton densities obtained from the CCFM evolution with a parton shower, following the uPDF. However, as reported in previous sections, the agreement of predictions obtained from CASCADE with measurements is, in general, not satisfactory, only in dedicated phase space regions, the agreement is rather good. One reason is that only gluon initiated processes were simulated, which is appropriate in the asymptotic region of applicability of BFKL, but is clearly not sufficient for the kinematic region at HERA.

A significant step forward has been made recently with the development and determination of the so-called Parton-Branching TMD distributions~\cite{Martinez:2018jxt,Hautmann:2017fcj}, which are uPDFs for all flavors and obtained with NLO evolution. These uPDFs can be applied in a new version of CASCADE~\cite{Baranov:2021uol} for detailed comparison with measurements, and impressive agreement with measurements at the LHC~\cite{Martinez:2019mwt} have been obtained. The new TMD (or uPDF) distributions are a significant extension of the early HERA unintegrated gluon densities.
With this development, a new and promising attempt to understand small $x$ measurements from HERA has been started~\cite{Monfared:2019uaj}.
In the last years a major step forward has been achieved with the automated calculation package {\sc KaTie}~\cite{vanHameren:2019puc,vanHameren:2016kkz} for off-shell matrix elements for all flavors in $\kt$-factorization, as well as with PEGASUS~\cite{Lipatov:2019oxs}, both also applicable to deep-inelastic scattering.

\section{Outlook}

The study and investigations of small $x$ effects at HERA, which were triggered by the pioneering work of Lev, have led to a very deep understanding of parton dynamics at large energies. Although, a final confirmation for the need of BFKL resummation could only be given recently for inclusive deep inelastic cross section at small $Q^2$, the studies of final states in deep inelastic scattering, which aimed to confirm deviations from predictions of collinear factorization, were extremely important for the investigations of new processes at large energies.

Recently a first complete set of unintegrated (Transverse Momentum Dependent) parton densities including all flavors was determined and can be used in event simulation with its unique application to parton showers. This development has great impact on present measurements (not only at small $x$) at the LHC but also influences the preparation for measurements at new colliders like the LHeC, the EIC or FCC, and thus the basic ideas of Lev which influenced so much a whole generation of  physicists at HERA, are alive also in the studies for  future colliders.

\bibliographystyle{mybibstyle-new.bst}
\raggedright
\bibliography{/Users/jung/Bib/hannes-bib}

\end{document}